\documentclass[12pt,amsmath,amssymb,]{revtex4}
\usepackage{amsfonts}
\usepackage{amsmath}
\usepackage{amssymb}
\usepackage{array}
\usepackage{bm}
\usepackage{braket}
\usepackage{breqn}
\usepackage{color}
\usepackage{dcolumn}
\usepackage{epsfig}
\usepackage{epsf}
\usepackage{epstopdf}
\usepackage{float}
\usepackage{graphicx}
\usepackage{graphics}
\usepackage{harpoon}
\usepackage{indentfirst}
\usepackage{mathrsfs}
\usepackage{multirow}
\usepackage{ulem}

\newcommand {\sla}[1]{ #1 \!\!\!\!/}

\newcommand{\RM}[1]{\textrm{\uppercase\expandafter{\romannumeral#1}}}
\begin{document}

\title{Corrections to the forward-limit dispersion relations for $\gamma Z$-exchange contributions}

\author{
Qian-Qian Guo and Hai-Qing Zhou\protect\footnotemark[1] \protect\footnotetext[1]{Email address: zhouhq@seu.edu.cn} \\
School of Physics,
Southeast University, NanJing 211189, China}

\date{\today}

\begin{abstract}
The weak charge of the proton $Q_{\textrm{w}}$ is one of the most fundamental quantities in physics. It can be determined by measuring the parity asymmetry $A_{\textrm{PV}}$ in elastic $ep$ scattering, where the $\gamma Z$-exchange contributions are crucial. For the past fifteen years, dispersion relations (DRs) in the forward limit have been widely used as a model-independent method to estimate these contributions. In this work, we study corrections to these forward-limit DRs. We first estimate these corrections using pointlike interactions as an illustrative example. We then estimate the $\gamma Z$-exchange contributions for the upcoming P2 experiment through both direct calculation and the forward-limit DRs, within the framework of low-energy effective interactions. The results indicate that the correction to the forward-limit DR for $\Box_{\gamma Z}^{V}$ is around 47\% for the upcoming P2 experiment, which will significantly modify the extracted value of $Q_{\textrm{w}}$.
 \end{abstract}

\maketitle

%%%%%%%%%%%%%%%%%%%%%%%%%%%%%%%%%%%%%%%%%%%%%%%%%%%%%%%%

\section{Introduction}
The proton is one of the most fundamental particles in our world, and studies of its structure have been ongoing for nearly a century. However, our understanding of its structure is still limited due to the nonperturbative nature of the strong interaction. In the past two decades, experimental measurements of the proton's structure have greatly improved, including measurements of its electromagnetic form factors \cite{proton-EM-FF-Ex-2000-Jones,proton-EM-FF-Ex-2002-Gayou,proton-EM-FF-Ex-2004-Christy,proton-EM-FF-Ex-2006-Qattan, proton-EM-FF-Ex-2010-Bernauer,proton-EM-FF-Ex-2011-Zhan,proton-EM-FF-Ex-2011-Ron,proton-EM-FF-Ex-2020-BESIII}, strange form factor (FF) \cite{proton-strange-FF-Ex-Muller-1997,proton-strange-FF-Ex-Hasty-2000,proton-strange-FF-Ex-Spayde-2004,proton-strange-FF-Ex-HAPPEX,proton-strange-FF-Ex-A4,proton-strange-FF-Ex-G0}, weak charge $Q_{\textrm{w}}$ \cite{Ex-Qweak}, size \cite{proton-size-Ex-2010-Pohl,proton-size-Ex-2013-Antognini,proton-size-Ex-2018-Fleurbaey,proton-size-Ex-2019-Bezginov,proton-size-Ex-2019-Xiong}, and others.

Similar to electromagnetic charge, $Q_{\textrm{w}}$ reflects the strength of the weak interaction of the proton at low energies. As quarks are confined, $Q_{\textrm{w}}$ becomes one of the most fundamental charges that can be measured in the standard model. In experiments, the parity-violating elastic $ep$ scattering provides a clean method for determining $Q_{\textrm{w}}$, where the asymmetry $A_{\textrm{PV}}$ is measured.

In the low-energy limit, $Q_{\textrm{w}}$ is proportional to the asymmetry $A_{\textrm{PV}}$, which means that the accurate determination of $Q_{\textrm{w}}$ requires precise measurements and analysis of $A_{\textrm{PV}}$. Theoretical calculations for this purpose focus on accurately estimating the interference between the one-photon-exchange and $\gamma Z$-exchange diagram. In the literature, various methods have been used to estimate the $\gamma Z$-exchange contributions \cite{gamma-Z-exchange-Th-1984-Marciano,gamma-Z-exchange-Th-2003-Erler,gamma-Z-exchange-Th-2007-Zhou,
gamma-Z-exchange-Th-2008-Tjon,gamma-Z-exchange-Th-DR-Gorchtein-2009,gamma-Z-exchange-Th-DR-Sibirtsev-2010,gamma-Z-exchange-Th-DR-2011-Blunden,
gamma-Z-exchange-Th-DR-2011-Rislow,gamma-Z-exchange-Th-DR-2011-Gorchtein,gamma-Z-exchange-Th-DR-2013-Hall,gamma-Z-exchange-Th-DR-2019-Seng,
gamma-Z-exchange-low-erergy-2023-zhouhq}. Among these methods, the forward limit dispersion relations (DRs) are widely applied and accepted as a model independent method to estimate $\gamma Z$-exchange contributions directly at the experimental regions.

In this study, we would like to discuss the detailed corrections to the forward-limit DRs in the low-energy region. The paper is structured as follows: In Sec. II, we provide the basic formula. In Sec. III, we present our numerical corrections to the forward-limit DRs in the pointlike theory and the low-energy model. Additionally, we discuss the reasons for any observed differences.

\section{Basic Formula}

\begin{figure}[htbp]
\centering
\includegraphics[height=4.8cm]{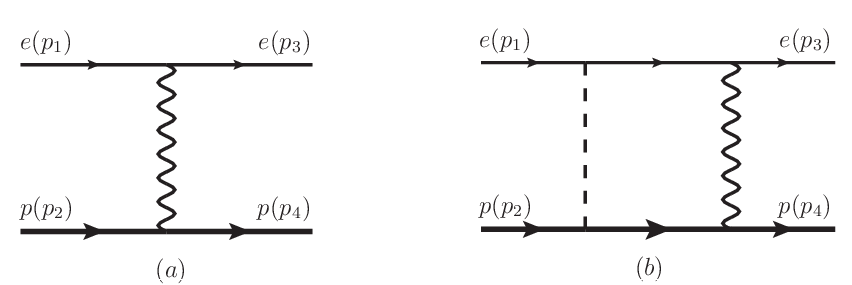}
\caption{Feynman diagrams for $ep\rightarrow ep$: ($a$) represents the one-photon-exchange diagram and ($b$) represents one of the $\gamma$Z-exchange diagrams.}
\label{Figure-general-gammaZ-exchange-in-ep}
\end{figure}

This asymmetry $A_{\textrm{PV}}$ in the parity-violating elastic $ep$ scattering is defined as follows:
\begin{equation}
\begin{aligned}
A_{\textrm{PV}} &\equiv&\frac{\sum\limits_{helicity}({\cal M}^{+}{\cal M}^{+*}-{\cal M}^{-}{\cal M}^{-*})}{\sum\limits_{helicity}({\cal M}^{+}{\cal M}^{+*}+{\cal M}^{-}{\cal M}^{-*})},
\end{aligned}
\end{equation}
where ${\cal M}^{+,-}$ are the helicity amplitudes in the laboratory frame with the incoming electron's helicity $\pm$, respectively. The corresponding one-photon-exchange and $\gamma Z$-exchange diagrams are depicted in Fig. \ref{Figure-general-gammaZ-exchange-in-ep}, where the interaction vertices between the electron and bosons are given by $\Gamma^{\mu}_{\gamma ee} = -ie\gamma^{\mu}$ and $\Gamma^{\mu}_{Zee} = \frac{ie}{4\sin\theta_{\textrm{w}}\cos\theta_{\textrm{w}}}[g_{e}^{V}\gamma^{\mu}+g_{e}^{A}\gamma^{\mu}\gamma_{5}]$ in the standard model, with $\theta_{\textrm{w}}$ the Weinberg angle.

According to the types of the interference, the $\gamma Z$-exchange contributions to $A_{\textrm{PV}}$ can be separated as follows:
\begin{equation}
A_{\textrm{PV}}^{\gamma Z}(E,Q^2) \equiv \frac{G_Ft}{4\sqrt{2}\pi\alpha_{e}}[\textrm{Re}[\Box_{\gamma Z}^{A}(E,Q^2)]+\textrm{Re}[\Box_{\gamma Z}^{V}(E,Q^2)]],
\end{equation}
where $G_F=\pi\alpha_e/(\sqrt{2}M_Z^2\sin^2\theta_{\textrm{w}}\cos^2\theta_{\textrm{w}})$ is the Fermi constant, $\alpha_e=e^2/4\pi$ is the fine structure constant,  $t=-Q^2=q^2=(p_4-p_2)^2=(p_1-p_3)^2$, and $E$ is the energy of incoming electron in laboratory frame with $(p_1+p_2)^2=M_N^2+2M_NE$ and $M_N$ the mass of the proton; $p_{1,2,3,4}$ are the momenta of the incoming electron, the incoming proton, the outgoing electron, and the outgoing proton, respectively. $\Box_{\gamma Z}^{A}(E,Q^2)$ and $\Box_{\gamma Z}^{V}(E,Q^2)$ are proportional to $g_{e}^{V}$ and $g_{e}^{A}$, respectively.

In the literature, the forward-limit DRs usually are used to estimate $\Box_{\gamma Z}^{V, A}(E,Q^2)$, which can be written as \cite{gamma-Z-exchange-Th-DR-Gorchtein-2009,gamma-Z-exchange-Th-DR-Sibirtsev-2010,gamma-Z-exchange-Th-DR-2011-Blunden,
gamma-Z-exchange-Th-DR-2011-Rislow,gamma-Z-exchange-Th-DR-2011-Gorchtein,gamma-Z-exchange-Th-DR-2013-Hall,gamma-Z-exchange-Th-DR-2019-Seng}
\begin{equation}
\begin{aligned}
\textrm{Re}[{\Box}_{\gamma Z}^{V}(E,Q^2)] \approx \textrm{Re}[{\Box}_{\gamma Z}^{V}(E,0)] = \frac{2 E}{\pi} P\Big[\int_{0}^{\infty} \frac{\textrm{Im}\big[{\Box}_{\gamma Z}^{V}(\bar{E}^{+},0)\big]}{\bar{E}^{2}-E^{2}} d \bar{E}\Big],   %\tag{$^{1st}$DR} \\
\\
\textrm{Re}[{\Box}_{\gamma Z}^{A}(E,Q^2)] \approx  \textrm{Re}[{\Box}_{\gamma Z}^{A}(E,0)] = \frac{2}{\pi} P\Big[\int_{0}^{\infty} \frac{\bar{E}\textrm{Im}\big[{\Box}_{\gamma Z}^{A}(\bar{E}^{+},0)\big]}{\bar{E}^{2}-E^{2}} d \bar{E}\Big], %\tag{$^{1st}$DR}
\end{aligned}
\label{eq:DR1}
\end{equation}
where $P$ refers to the principle value integration, and $\bar{E}^{+}=\bar{E}+i0^+$. Naively, one question is, how much is the difference between $\textrm{Re}[{\Box}_{\gamma Z}^{V,A}(E,Q^2)]$ and $\textrm{Re}[{\Box}_{\gamma Z}^{V,A}(E,0)]$ in the low-energy region? In Ref. \cite{Gorchtein-2011-PRC}, $\textrm{Re}[{\Box}_{\gamma Z}^{V,A}(E,Q^2)]$ is estimated using the following continued formula:
 \begin{equation}
\square_{\gamma Z}(E, t)\approx \square_{\gamma Z}(E, 0) \frac{\exp (-B|t| / 2)}{F_1^{\gamma p}(t)},
\label{eq:Box-continue}
\end{equation}
with $B$ some parameter.

Another naive approximation is to use the following expressions:
 \begin{equation}
\begin{aligned}
\textrm{Re}[{\Box}_{\gamma Z}^{V}(E,Q^2)] \approx {\cal C}_{\gamma Z}^{V}(E,Q^2) \equiv \frac{2 E}{\pi} P\Big[\int_{0}^{\infty} \frac{\textrm{Im}[{\Box}_{\gamma Z}^{V}(\bar{E}^{+},Q^2)\big]}{\bar{E}^{2}-E^{2}} d \bar{E}\Big],
\\
\textrm{Re}[{\Box}_{\gamma Z}^{A}(E,Q^2)] \approx {\cal C}_{\gamma Z}^{A}(E,Q^2) \equiv \frac{2 }{\pi} P\Big[\int_{0}^{\infty} \frac{\bar{E}\textrm{Im}\big[{\Box}_{\gamma Z}^{A}(\bar{E}^{+},Q^2)\big]}{\bar{E}^{2}-E^{2}} d \bar{E}\Big]. %\tag{$^{1st}$DR}
\end{aligned}
\label{eq:DR2}
\end{equation}
We would like to point out that, at finite $Q^2$, the approximation in Eq. (\ref{eq:DR2}) slightly differs from the following approximation:
\begin{equation}
\begin{aligned}
\textrm{Re}[{\Box}_{\gamma Z}^{V}(E,Q^2)] \approx {\cal D}_{\gamma Z}^{V}(E,Q^2) \equiv \frac{2 \nu}{\pi} P\Big[\int_{\nu_{th}}^{\infty} \frac{\textrm{Im}[{\Box}_{\gamma Z}^{V}(\bar{E}^{+},Q^2)\big]}{\bar{\nu}^{2}-\nu^{2}} d \bar{\nu}\Big],
\\
\textrm{Re}[{\Box}_{\gamma Z}^{A}(E,Q^2)] \approx {\cal D}_{\gamma Z}^{A}(E,Q^2) \equiv \frac{2 }{\pi} P\Big[\int_{\nu_{th}}^{\infty} \frac{\bar{\nu}\textrm{Im}[{\Box}_{\gamma Z}^{A}(\bar{E}^{+},Q^2)\big]}{\bar{\nu}^{2}-\nu^{2}} d \bar{\nu}\Big], %\tag{$^{1st}$DR}
\end{aligned}
\label{eq:DR3}
\end{equation}
where $\nu \equiv 2(p_1+p_2)^2-2M_{N}^2-Q^2=4M_NE-Q^2$ and $\nu_{th}=-Q^2$.

When $Q^2=0$, one has
\begin{equation}
{\cal D}_{\gamma Z}^{V,A}(E,0)={\cal C}_{\gamma Z}^{V,A}(E,0)=\textrm{Re}[{\Box}_{\gamma Z}^{V,A}(E,0)].
\end{equation}

Before going to discuss the difference between these approximations, we review some basic properties of $A_{\textrm{PV}}$ \cite{gamma-Z-exchange-low-erergy-2023-zhouhq}. The full $\gamma Z$-exchange amplitude can be separated into a parity-conserved (PC) part and a parity-violated (PV) part as
\begin{equation}\begin{aligned}
{\cal M}_{\gamma Z} &\equiv  {\cal M}^{\textrm{PC}}_{\gamma Z}+{\cal M}^{\textrm{PV}}_{\gamma Z}, \\
{\cal M}^{\textrm{PV}}_{\gamma Z} &\equiv  g_{e}^A {\cal M}^{V}_{\gamma Z}+ g_{e}^V {\cal M}^{A}_{\gamma Z}.
\label{eq-separate-V-A}
\end{aligned}\end{equation}
After taking the approximation $m_e=0$ with $m_e$ the mass of electron, the amplitudes ${\cal M}^{V,A}_{\gamma Z}$ can be written as
\begin{equation}\begin{aligned}
{\cal M}^{V}_{\gamma Z} \equiv \sum_{i=1}^{3} {\cal F}_{\gamma Z,i}^{V}{\cal P}_{i}^{V},
{\cal M}^{A}_{\gamma Z}\equiv \sum_{i=1}^{3} {\cal F}_{\gamma Z,i}^{A}\mathcal{P}_{i}^{A},
\label{eq-amplitude-total}
\end{aligned}\end{equation}
where the general invariant amplitudes ${\cal P}_{i}^{V}$ and ${\cal P}_{i}^A$ are chosen as \cite{gamma-Z-exchange-low-erergy-2023-zhouhq}
\begin{equation}\begin{aligned}
{\cal P}_{1}^{V} &\equiv \left[\bar{u}_{3} \gamma_{\mu} \gamma_{5} u_{1}\right]\left[\bar{u}_{4} \gamma^{\mu} u_{2}\right], \\
{\cal P}_{2}^{V} &\equiv \frac{1}{Q}\left[\bar{u}_{3} \gamma_{\mu} \gamma_{5} u_{1}\right]\left[\bar{u}_{4} i\sigma^{\mu\nu}q_\nu u_{2}\right],  \\
{\cal P}_{3}^{V} &\equiv \frac{1}{M_NQ}\left[\bar{u}_{3}\sla{P}\gamma_{5} u_{1}\right]\left[\bar{u}_{4}\sla{K}u_{2}\right], \\
{\cal P}_{1}^{A} &\equiv \left[\bar{u}_{3} \gamma^{\mu}  u_{1}\right]\left[\bar{u}_{4} \gamma_{\mu}\gamma_{5} u_{2}\right],  \\
{\cal P}_{2}^{A} &\equiv \frac{1}{Q}\left[\bar{u}_{3} \gamma^{\mu} u_{1}\right]\left[\bar{u}_{4} \gamma_{\mu} \sla{K} \gamma_{5} u_{2}\right],  \\
{\cal P}_{3}^{A} &\equiv \frac{1}{M_NQ}\left[\bar{u}_{3}\sla{P} u_{1}\right]\left[\bar{u}_{4} \sla{K}  \gamma_{5}u_{2}\right],
\label{eq-invariant-amplitudes-reference}
\end{aligned}\end{equation}
with $P=p_2+p_4$, $K=p_1+p_3$.

After some calculations, $A_{\textrm{PV}}^{\gamma Z}$ can be expressed as
\begin{equation}\begin{aligned}
A_{\textrm{PV}}^{\gamma Z}
&=&\frac{1}{e^2\sigma}\Big(g_e^{A}\sum_{i=1}^{3}{\cal N}_{i}^{V}\textrm{Re}[{\cal F}_{\gamma Z,i}^{V}] +g_e^{V}\sum_{i=1}^{3}{\cal N}_{i}^{A}\textrm{Re}[{\cal F}_{\gamma Z,i}^{A}]\Big),
\label{eq:APV}
\end{aligned}\end{equation}
with
\begin{equation}\begin{aligned}
{\cal N}_{1}^{V}&=8M_N^2Q^2[(\nu^2-4M_N^2Q^2+Q^4)F_1+2Q^4F_2],\\
{\cal N}_{2}^{V}&=4M_NQ[8M_N^2Q^4F_1+Q^2(\nu^2+4M_N^2Q^2-Q^4)F_2],\\
{\cal N}_{3}^{V}&=8M_NQ\nu(\nu^2-4M_N^2Q^2-Q^4)F_1,\\
{\cal N}_{1}^{A}&=16M_N^2Q^4\nu(F_1+F_2),\\
{\cal N}_{2}^{A}&=4M_NQ^3[8M_N^2Q^2F_1+(\nu^2+4M_N^2Q^2-Q^4)F_2],\\
{\cal N}_{3}^{A}&=8M_NQ^3(\nu^2-4M_N^2Q^2-Q^4)(F_1+F_2),
\label{Eq-expressions-of-N}
\end{aligned}\end{equation}
and
\begin{equation}\begin{aligned}
\sigma = 4F_1^2M_N^2(\nu^2-4M_N^2Q^2+Q^4)+16F_1F_2M_N^2Q^4+F_2^2Q^2(\nu^2+4M_N^2Q^2-Q^4),
\end{aligned}\end{equation}
where $F_{1,2}$ are the electromagnetic FFs of proton.

To discuss the details of the differences between the above approximations, we take two types of interactions as examples to illustrate their properties. In the first case, we treat the proton as a pointlike particle, where the corresponding interaction vertices can be well defined and expressed as
\begin{equation}
\begin{aligned}
\overline{\Gamma}^{\mu}_{\gamma pp} &=  ie \gamma^{\mu},  \\
\overline{\Gamma}^{\mu}_{Zpp} &= \frac{ie}{4\sin\theta_{\textrm{w}}\cos\theta_{\textrm{w}}}[g_{1}\gamma^{\mu}+g_{3}\gamma^{\mu}\gamma^{5}],
\label{eq-vertex}
\end{aligned}
\end{equation}
which also represent the leading-order (LO) low-energy interactions.  In the second case, we consider the $\gamma pp$ and $Zpp$ interactions at the LO and the next-to-leading order (NLO) of momentum. These interactions can be described as follows:
\begin{equation}
\begin{aligned}
\Gamma^{\mu}_{\gamma pp} =&  ie[F_{1}\gamma^{\mu}+F_{2}\frac{i\sigma^{\mu\nu}}{2M_{N}}q_{\nu}],  \\
\Gamma^{\mu}_{Zpp} =& \frac{ie}{4\sin\theta_{\textrm{w}}\cos\theta_{\textrm{w}}}[g_{1}\gamma^{\mu}+g_{2}\frac{i\sigma^{\mu\nu}}{2M_{N}}q_{\nu}+g_{3}\gamma^{\mu}\gamma^{5}].
\label{eq-vertex}
\end{aligned}
\end{equation}

Through these interactions, the contributions of $\gamma Z$ exchange can be directly calculated, and the forward-limit dispersion relations (DRs) can also be examined within the energy regions where these interactions are applicable. In practical calculations, the package FeynCalc10.0 \cite{FenyCalc} is used to deal with Dirac matrix, the package PackageX3.0 \cite{PacakgeX} is used to do the loop integration, and the package LoopTools \cite{LoopTools} is used for cross-check.

\section{Results and Discussion}

In the pointlike interaction case, the direct calculation  shows that ${\cal F}_{\gamma Z,i}^{V,A}(E,Q^2)$ satisfy DRs such as Eq. (\ref{eq:DR3}) exactly for any $Q^2$ \footnotemark[2]\footnotetext[2]{${\cal F}_{\gamma Z,1}^{V,A}(E,Q^2)$ satisfy the 1$^{st}$ DR and ${\cal F}_{\gamma Z,3}^{V,A}(E,Q^2)$ satisfy the 2$^{nd}$ DR in Eq. (\ref{eq:DR3}).}. This means the results for $\textrm{Re}[\Box_{\gamma Z}^{V,A}(E,Q^2)]$ obtained through the direct calculation  are identical to those obtained by first dispersing ${\cal F}_{\gamma Z,i}^{V,A}(E,Q^2)$ and then substituting into Eq. (\ref{eq:APV}). However, at finite $Q^2$, $\Box_{\gamma Z}^{V,A}(E,Q^2)$ do not satisfy the similar DRs. The reason  can be traced to the double pole in Eq. (\ref{eq:APV}). This double pole gives rise to the following DRs:
\begin{equation}\begin{aligned}
\textrm{Re}\Big[{\Box}_{\gamma Z}^{V}(E,Q^{2})\Big] =\frac{c_V\nu}{\nu^2-\nu_p^2}+\frac{2 \nu}{\pi} P\Big[\int_{\nu_{th}}^{\infty} \frac{\textrm{Im}\big[{\Box}_{\gamma Z}^{V}(Q^{2}, \bar{\nu}^{+})\big]}{\bar{\nu}^{2}-\nu^{2}} d \bar{\nu}\Big], \\
\textrm{Re}\Big[{\Box}_{\gamma Z}^{A}(E,Q^{2})\Big] =\frac{c_A}{\nu^2-\nu_p^2}+\frac{2 }{\pi} P\Big[\int_{\nu_{th}}^{\infty} \frac{\bar{\nu}\textrm{Im}\big[{\Box}_{\gamma Z}^{A}(Q^{2}, \bar{\nu}^{+})\big]}{\bar{\nu}^{2}-\nu^{2}} d \bar{\nu}\Big],
\label{eq:DR-3}
\end{aligned}\end{equation}
where $\nu_p$ is the zero point of $\sigma$, and $c_{A,V}$ are constants which are only dependent on $Q$.

\begin{figure}[htbp]
\centering
\includegraphics[height=10.5cm]{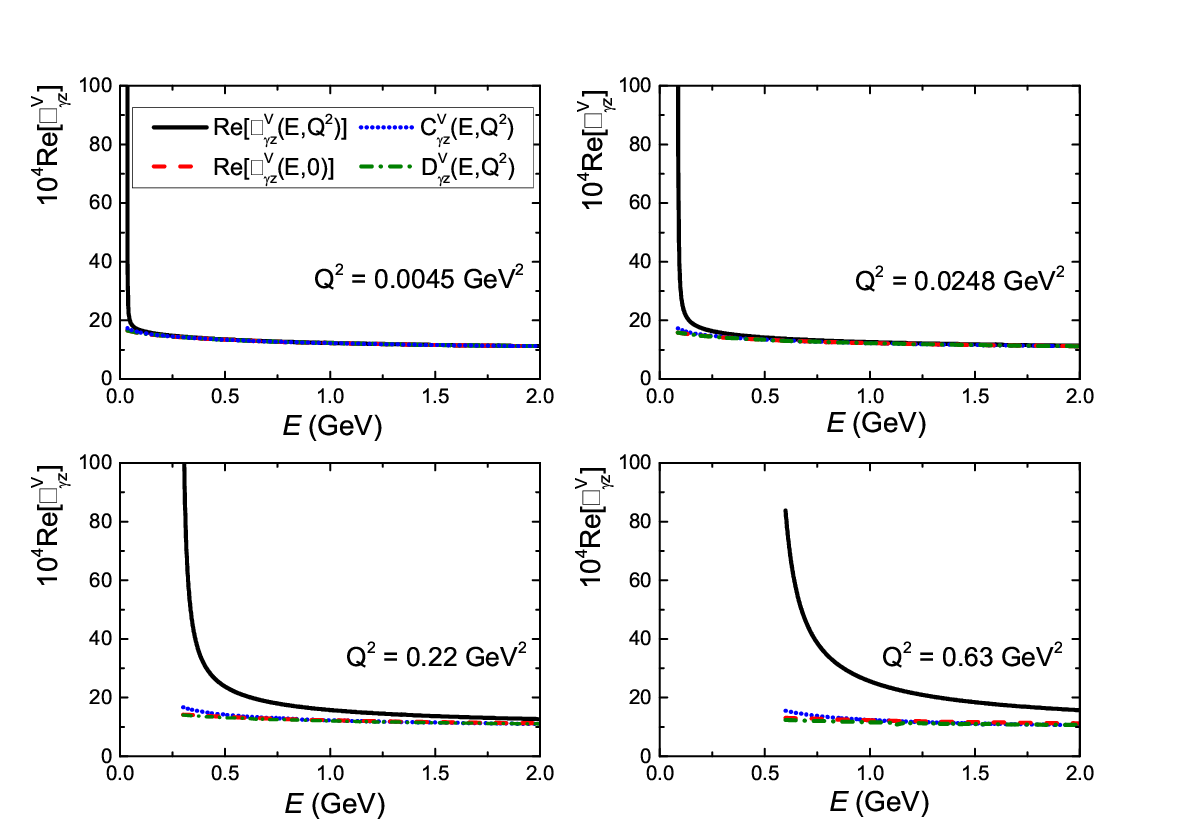}
\caption{Numerical results for $\textrm{Re}[\Box_{\gamma Z}^{V}(E,Q^2)]$, $\textrm{Re}[\Box_{\gamma Z}^{V}(E,0)]$, ${\cal C}_{\gamma Z}^{V}(E,Q^2)$, and ${\cal D}_{\gamma Z}^{V}(E,Q^2)$  in the physical region with $E\geq E_{min}\equiv Q(\sqrt{4M_N^2+Q^2}+Q)/4M_N$, where the pointlike interactions are considered. The solid-black, dashed-red, dotted-blue, and dashed-dotted-olive curves correspond to $\textrm{Re}[\Box_{\gamma Z}^{V}(E,Q^2)]$, $\textrm{Re}[\Box_{\gamma Z}^{V}(E,0)]$, ${\cal C}_{\gamma Z}^{V}(E,Q^2)$, and ${\cal D}_{\gamma Z}^{V}(E,Q^2)$, respectively.}
\label{Figure-V11-comparison}
\end{figure}

To quantify the differences, we present the numerical results for $\textrm{Re}[\Box_{\gamma Z}^{V}(E,Q^2)]$, $\textrm{Re}[\Box_{\gamma Z}^{V}(E,0)]$ , ${\cal C}_{\gamma Z}^{V}(E,Q^2)$, and ${\cal D}_{\gamma Z}^{V}(E,Q^2)$ in Fig. \ref{Figure-V11-comparison}, where the parameters $g_{e}^{V}=-0.076$, $g_{e}^{A}=1$, $g_1=0.076$, and $g_3=-0.95$ are chosen \cite{gamma-Z-exchange-Th-2007-Zhou}, and the energy $E$ is restricted to the physical region $E\geq E_{min}\equiv Q(\sqrt{4M_N^2+Q^2}+Q)/4M_N$ at the corresponding  experimental $Q^2$ \cite{proton-strange-FF-Ex-Muller-1997,proton-strange-FF-Ex-Hasty-2000,
proton-strange-FF-Ex-Spayde-2004,proton-strange-FF-Ex-HAPPEX,proton-strange-FF-Ex-A4,proton-strange-FF-Ex-G0,Ex-Qweak,P2-experiment-2018}. The differences between $\textrm{Re}[\Box_{\gamma Z}^{A}(E,Q^2)]$, $\textrm{Re}[\Box_{\gamma Z}^{A}(E,0)]$, ${\cal C}_{\gamma Z}^{A}(E,Q^2)$, and ${\cal D}_{\gamma Z}^{A}(E,Q^2)$ are small and thus are not presented. The comparisons clearly indicate that $\textrm{Re}[\Box_{\gamma Z}^{V}(E,0)]$, ${\cal C}_{\gamma Z}^{V}(E,Q^2)$, and ${\cal D}_{\gamma Z}^{V}(E,Q^2)$ are very similar to each other in almost the entire range. However, for very small values of $E$ or $Q^2>0.22$ GeV$^2$, there are significant differences between these quantities and $\textrm{Re}[\Box_{\gamma Z}^{V}(E,Q^2)]$. Additionally, the results reveal that $\textrm{Re}[\Box_{\gamma Z}^{V}(E,Q^2)]/\textrm{Re}[\Box_{\gamma Z}^{V}(E,0)]$ not only depends on $Q^2$ but also exhibits a strong dependence on $E$, particularly when $E$ is small. This finding suggests that the simple continuity equation given by Eq. (\ref{eq:Box-continue}) is not applicable when dealing with small values of $E$.

Detailed numerical comparisons in the corresponding  experimental energy regions \cite{proton-strange-FF-Ex-Muller-1997,proton-strange-FF-Ex-Hasty-2000,
proton-strange-FF-Ex-Spayde-2004,proton-strange-FF-Ex-HAPPEX,proton-strange-FF-Ex-A4,proton-strange-FF-Ex-G0,Ex-Qweak,P2-experiment-2018}, are provided in Table \ref{Tab-comparison-at-experimental-points}. These comparisons clearly indicate that the corrections are correlated with the values of $(E-E_{min})M_N/Q^2$.

\begin{table}[htbp]
\renewcommand\arraystretch{1.5}
\centering
\begin{tabular}{p{3.5cm}<{\centering}p{1.5cm}<{\centering}p{1.5cm}<{\centering}p{1.5cm}<{\centering}
p{1.5cm}<{\centering}p{1.5cm}<{\centering}p{1.5cm}<{\centering}p{1.5cm}<{\centering}}
\hline
\hline
 & & & & {Experiment} \\
\cline{2-8}
 &P2 &Qweak &G0 & G0&HAPPEX  &A4 &SAMPLE \\
\hline
$Q^2$ (GeV$^2$) &0.0045 &0.0248 & 0.22 &0.63 &0.624 &0.23&0.1\\
\hline
$E$ (GeV)& 0.155& 1.15&0.362 &0.687 &3.48 &0.854&0.2 \\
\hline
$(E-E_{min})M_N/Q^2$& 25& 432  & 0.26 & 0.13  & 4.34 & 2.22 & 0.12\\
\hline
\hline
$\frac{{\cal C}_{\gamma Z}^{V}(E,0)}{\textrm{Re}[\Box_{\gamma Z}^{V}(E,Q^2)]}$ &96.6\% &97.6\% &35.8\% &27.2\% & 83.2\% &73.3\% &25.0\% \\
\hline
$\frac{{\cal C}_{\gamma Z}^{V}(E,Q^2)}{\textrm{Re}[\Box_{\gamma Z}^{V}(E,Q^2)]}$ &96.6\% &97.4\% &35.3\% &25.7\% & 79.7\% &71.9\% &24.9\% \\
\hline
$\frac{{\cal D}_{\gamma Z}^{V}(E,Q^2)}{\textrm{Re}[\Box_{\gamma Z}^{V}(E,Q^2)]}$ &97.2\% &97.5\% &40.1\% &30.2\% &78.6\%  &73.8\% &28.2\%\\
\hline
\hline
$\frac{{\cal C}_{\gamma Z}^{A}(E,0)}{\textrm{Re}[\Box_{\gamma Z}^{A}(E,Q^2)]}$  & 99.1\% & 99.4\%  & 65.1\% &56.9\% &95.1\% &91.7\%  &49.6\%\\
\hline
$\frac{{\cal C}_{\gamma Z}^{A}(E,Q^2)}{\textrm{Re}[\Box_{\gamma Z}^{A}(E,Q^2)]}$  & 99.7\% & 99.9\%  & 71.0\% &64.8\% &98.7\% &96.4\%  &53.2\%\\
\hline
$\frac{{\cal D}_{\gamma Z}^{A}(E,Q^2)}{\textrm{Re}[\Box_{\gamma Z}^{A}(E,Q^2)]}$ &99.8\% &99.9\% &71.6\% &64.8\% &99.1\% &96.8\% &53.6\%\\
\hline
\end{tabular}
\caption{Comparisons of $\textrm{Re}[\Box_{\gamma Z}^{V}(E,Q^2)]$, $\textrm{Re}[\Box_{\gamma Z}^{V}(E,0)]$, ${\cal C}_{\gamma Z}^{A}(E,Q^2)$, and ${\cal D}_{\gamma Z}^{A}(E,Q^2)$ at various experimental energy points, where the pointlike interactions are considered.}
\label{Tab-comparison-at-experimental-points}
\end{table}

The above results are obtained exactly within the pointlike particle approximation. For the physical $ep$ scattering, an interesting property is that the mass-center energy of the coming P2 experiment \cite{P2-experiment-2018} is below the resonance $\Delta(1232)$, where we can expect that the low-energy effective interactions in the second case can be used to estimate the $\gamma Z$-exchange contributions.

Within this framework of LO and NLO interactions, the coefficients ${\cal F}_{\gamma Z,i}^{V,A}(E,Q^2)$ still satisfy similar DRs such as Eq. (\ref{eq:DR3}) exactly, except for terms proportional to $F_2g_2$ whose real parts contain UV divergences and satisfy once-subtracted DRs. In the effective theory, the presence of UV divergences implies that some contact terms need to be introduced to absorb these divergences. These contact terms correspond to the subtracted terms in the subtracted DRs. Since the contributions from $F_2g_2$ in ${\cal F}_{\gamma Z,i}^{V}$ are of higher order in $Q$ or $E$, we neglect them in the current analysis.

As anticipated, when $Q^2\rightarrow0$ and $(E-E_{min})M_N/Q^2>10^4$, $\textrm{Re}[\Box_{\gamma Z}^{V,A}(E,Q^2)]$, $\textrm{Re}[\Box_{\gamma Z}^{V,A}(E,0)]$, ${\cal C}_{\gamma Z}^{V,A}(E,Q^2)$, and ${\cal D}_{\gamma Z}^{V,A}(E,Q^2)$ exhibit nearly identical behavior, resembling the case of pointlike interactions.

For the energy point of the upcoming P2 experiment, where $Q^{2}=0.0045$ GeV$^2$ and $E=0.155$ GeV, we obtain the following results \cite{gamma-Z-exchange-low-erergy-2023-zhouhq}:
\begin{equation}
\begin{aligned}
\textrm{Re}[\Box_{\gamma Z}^{V}(\textrm{P2})]=&10^{-4}\frac{g_e^{A}}{\sigma}(230.269F_1^2g_1+7.582F_1^2g_2+13.928 F_1F_2g_1\\
&~~~~~~~~~~~+4.090F_2^2g_1+4.394F_2F_1g_2), \\
\textrm{Re}[\Box_{\gamma Z}^{A}(\textrm{P2})] =&10^{-4}\frac{g_e^{V}}{\sigma}(410.700F_1^2+410.961F_1F_2-0.115F_2^2)g_3.
\end{aligned}
\end{equation}
The uncertainties in the estimations of $\textrm{Re}[\Box_{\gamma Z}^{V,A}(\textrm{P2})]$ are therefore linked to uncertainties in the low-energy coupling constants and the corrections from higher orders. By taking the low-energy coupling constants as $F_1=1$, $F_2=1.793$, $g_1=0.076$, $g_2=2.08$, and $g_3=-0.95$ \cite{gamma-Z-exchange-Th-2007-Zhou}, we find that $\textrm{Re}[\Box_{\gamma Z}^{V,A}(0,0.155 \textrm{GeV})]$, as well as ${\cal C}_{\gamma Z}^{V,A}(\textrm{P2})$ and ${\cal D}_{\gamma Z}^{V,A}(\textrm{P2})$, remain relatively similar to each other.  Their differences compared to the full results $\textrm{Re}[\Box_{\gamma Z}^{V,A}(\textrm{P2})]$ are as follows:
\begin{equation}
\begin{aligned}
\frac{{\cal C}_{\gamma Z}^{V}(\textrm{P2})=0.002221}{\textrm{Re}[\Box_{\gamma Z}^{V}(\textrm{P2})]=0.004685}&=47.41\%,\\
\frac{{\cal C}_{\gamma Z}^{A}(\textrm{P2})=0.007370}{\textrm{Re}[\Box_{\gamma Z}^{A}(\textrm{P2})]=0.007383}&=99.82\%.
\end{aligned}
\end{equation}
These comparisons reveal an important property when the physical interactions are considered: for the upcoming P2 experiment, the physical $\textrm{Re}[\Box_{\gamma Z}^{V}(E,Q^2)]$ is significantly larger than the forward-limit $\textrm{Re}[\Box_{\gamma Z}^{V}(E,0)]$ or ${\cal C}_{\gamma Z}^{V}(E,Q^2)$, while $\textrm{Re}[\Box_{\gamma Z}^{A}(E,Q^2)]$ is very close to  forward-limit $\textrm{Re}[\Box_{\gamma Z}^{A}(E,0)]$ or ${\cal C}_{\gamma Z}^{A}(E,Q^2)$.

\begin{figure}[htbp]
\centering
\includegraphics[height=8.5cm]{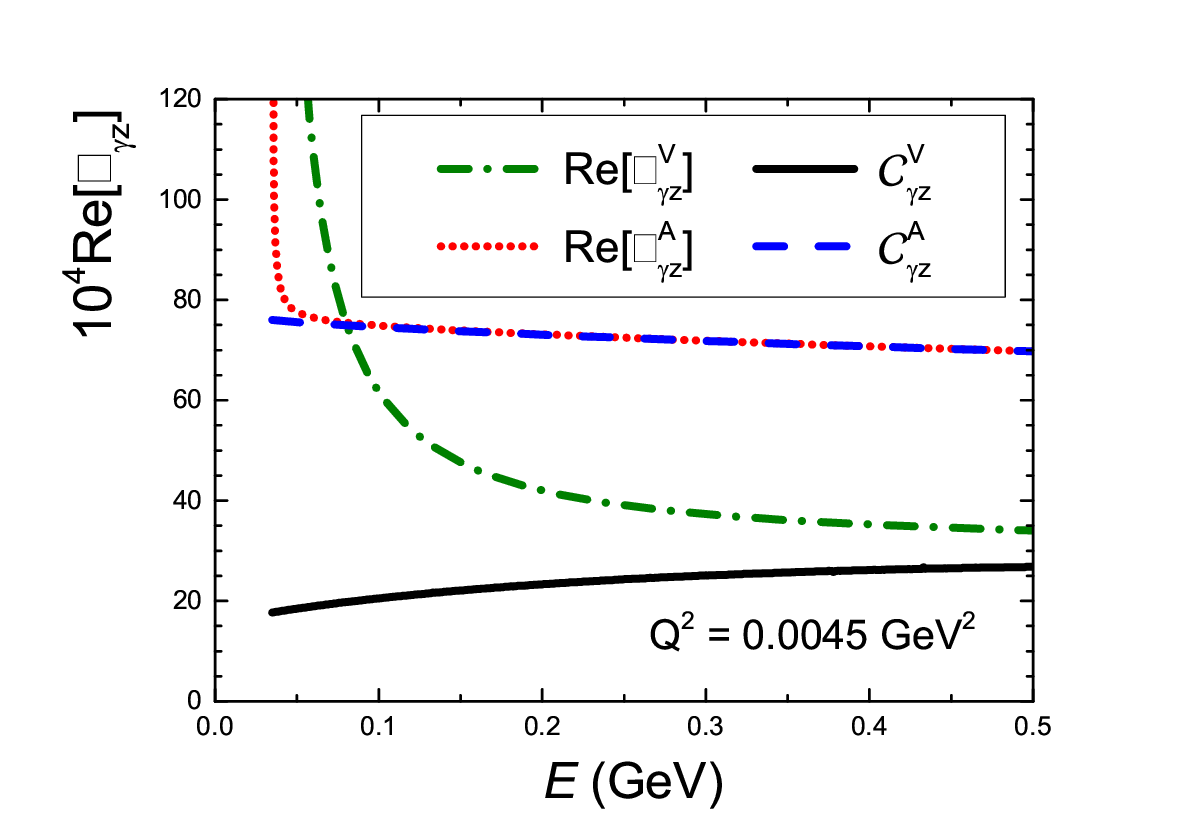}
\caption{Comparison of $\textrm{Re}[\Box_{\gamma Z}^{V,A}(E,Q^2)]$ and ${\cal C}_{\gamma Z}^{V,A}(E,Q^2)$ in the region with $E\geq E_{min}$, where the LO+NLO low-energy interactions are considered. The dashed-dotted olive, solid black, dotted red, and dashed blue curves correspond to $\textrm{Re}[\Box_{\gamma Z}^{V}(E,Q^2)]$, ${\cal C}_{\gamma Z}^{V}(E,Q^2)$, $\textrm{Re}[\Box_{\gamma Z}^{A}(E,Q^2)]$, and ${\cal C}_{\gamma Z}^{A}(E,Q^2)$, respectively.}
\label{Figure-Box-VA-NLO-interactions}
\end{figure}

In Fig.~\ref{Figure-Box-VA-NLO-interactions}, we present the $E$ dependence of $\textrm{Re}[\Box_{\gamma Z}^{V,A}(E,Q^2)]$  and ${\cal C}_{\gamma Z}^{V,A}(E,Q^2)$ at $Q^2=0.0045$ GeV$^2$ with LO+NLO interactions. The result for $\textrm{Re}[\Box_{\gamma Z}^{A}(E,Q^2)]$ at small $E$ such as $0.05$ GeV is consistent with those reported in Refs. \cite{gamma-Z-exchange-Th-DR-2011-Blunden,gamma-Z-exchange-Th-DR-2019-Seng}, but the behavior of $\textrm{Re}[\Box_{\gamma Z}^{V}(E,Q^2)]$ at small physical $E$ is much different from those reported in Ref. \cite{gamma-Z-exchange-Th-DR-2011-Rislow,gamma-Z-exchange-Th-DR-2011-Gorchtein,gamma-Z-exchange-Th-DR-2013-Hall}.

Further analysis reveals that the larger corrections in $\textrm{Re}[\Box_{\gamma Z}^{V}(E,Q^2)]$ and the significantly different behavior of $\textrm{Re}[\Box_{\gamma Z}^{V}(E,Q^2)]$ from the references are associated with three reasons. First, the forward limit is only accurate when $(E-E_{min})M_N/Q^2\rightarrow \infty$ for $Q<M_N$, which is not a good approximation for the P2 experiment. Second, the magnitude of the ratio $g_2/g_1$ is relatively large, which plays a significant role in contributing to the observed large corrections. Third, the nonzero $F_2$ also gives considerable corrections even $g_2$ is taken as zero.

\begin{figure}[htbp]
\centering
\includegraphics[height=8.5cm]{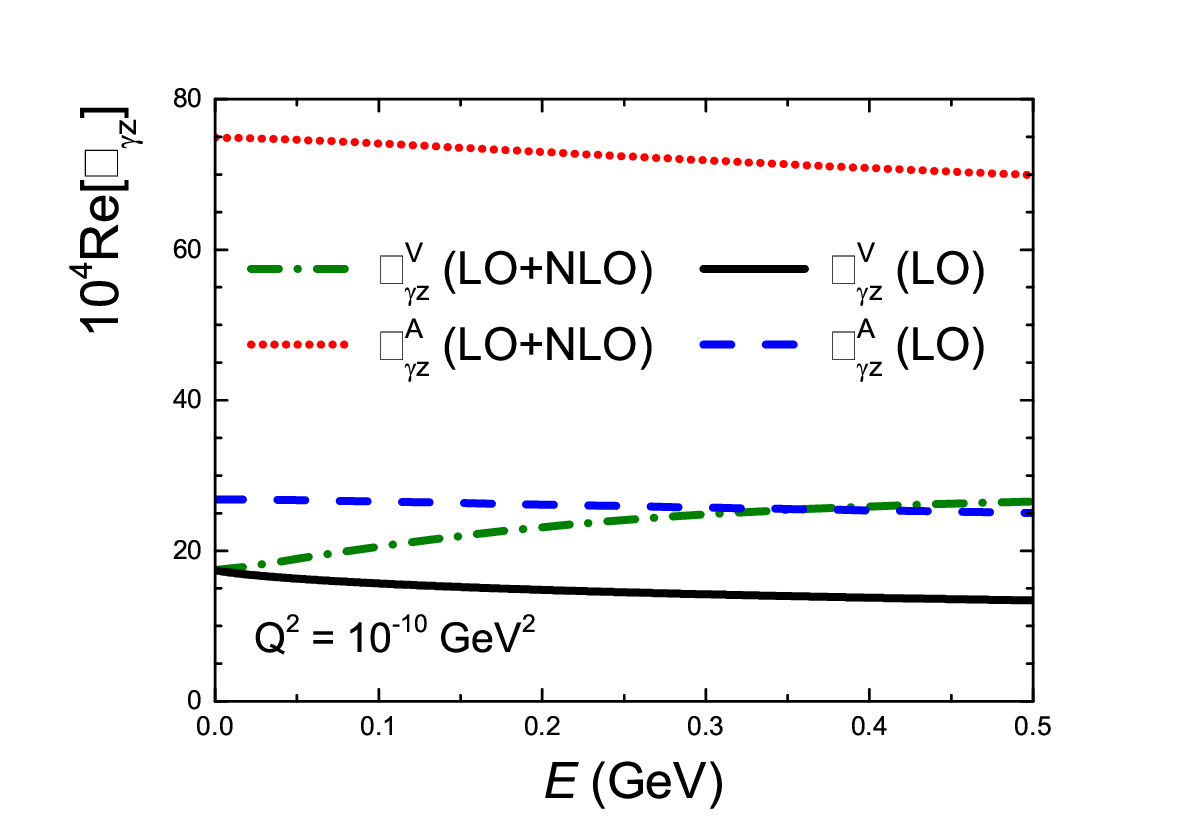}
\caption{$\textrm{Re}[\Box_{\gamma Z}^{V,A}(E,Q^2)]$ obtained with the LO and LO+NLO low-energy interactions, respectively. The dashed-dotted olive and solid black curves correspond to $\textrm{Re}[\Box_{\gamma Z}^{V}(E,Q^2)]$ with LO+NLO and LO interactions, respectively. The dotted red and dashed blue curves correspond to $\textrm{Re}[\Box_{\gamma Z}^{A}(E,Q^2)]$  with LO+NLO and LO interactions, respectively.}
\label{Figure-Box-VA-different-interactions}
\end{figure}

In Fig. \ref{Figure-Box-VA-different-interactions}, we present $E$ dependence of $\textrm{Re}[\Box_{\gamma Z}^{V,A}(E,Q^2)]$ obtained using LO interactions and LO+NLO interactions as $Q^2\rightarrow 0$. The results indicate that, for very small values of $E$, the NLO interactions give a large contribution to $\textrm{Re}[\Box_{\gamma Z}^{A}(E,Q^2)]$ due to the nonzero $F_2$, but a very small contribution to $\textrm{Re}[\Box_{\gamma Z}^{V}(E,Q^2)]$.

In summary, the widely used forward-limit DRs for $\Box_{\gamma Z}^{V,A}(E,0)$ work well, as expected in the region with $(E-E_{min})M_N/Q^2\rightarrow \infty$ for $Q<M_N$. However, when $(E-E_{min})M_N/Q^2$ is not sufficiently large, contributions beyond the forward limit must be taken into account, and it is recommended to use DRs such as Eq. (\ref{eq:DR3}) to estimate the coefficients ${\cal F}_{\gamma Z,i}^{V,A}(E,Q^2)$ instead of $\Box_{\gamma Z}^{V,A}(E,Q^2)$. For the upcoming P2 experiment, the $\gamma Z$ contributions are estimated using the LO and NLO low-energy effective interactions, and the numerical results show that the forward-limit DRs used in the literature may potentially underestimate $\textrm{Re}[\Box_{\gamma Z}^{V}(\textrm{P2})]$ by as much as $47\%$.

\section{Acknowledgments}
H.Q.Z. would like to thank Shin-Nan Yang and Zhi-Hui Guo for their helpful suggestions and discussions. This work is funded in part by the National Natural Science Foundations of China under Grants No. 12075058, No. 12150013, and No. 11975075.


\begin{thebibliography}{99}
%\bibitem{proton-EM-FF-Ex-1994}
%L. Andivahis {\it et al.}, Phys. Rev. D {\bf  50}, 5491 (1994); R. C. Walker {\it et al.}, Phys. Rev. D {\bf  49}, 5671 (1994);
\bibitem{proton-EM-FF-Ex-2000-Jones}
M. K. Jones {\it et al.} (JLab Hall A Coll.), Phys. Rev. Lett. {\bf  84}, 1398 (2000).

\bibitem{proton-EM-FF-Ex-2002-Gayou}
O. Gayou {\it et al.} (JLab Hall A Coll.), Phys. Rev. Lett. {\bf 88}, 092301 (2002).

\bibitem{proton-EM-FF-Ex-2004-Christy}
M. E. Christy  {\it et al.}, Phys. Rev. C \textbf{70}, 015206 (2004).

\bibitem{proton-EM-FF-Ex-2006-Qattan}
I. A. Qattan {\it et al.}, Phys. Rev. Lett. \textbf{94}, 142301 (2005).

\bibitem{proton-EM-FF-Ex-2010-Bernauer}
%\cite{A1:2010nsl}
%\bibitem{A1:2010nsl}
J.~C.~Bernauer \textit{et al.} [A1],
%``High-precision determination of the electric and magnetic form factors of the proton,''
Phys. Rev. Lett. \textbf{105}, 242001 (2010).
%doi:10.1103/PhysRevLett.105.242001
%[arXiv:1007.5076 [nucl-ex]].
%436 citations counted in INSPIRE as of 21 Jun 2023

\bibitem{proton-EM-FF-Ex-2011-Zhan}
%\cite{Zhan:2011ji}
%\bibitem{Zhan:2011ji}
X.~Zhan, \textit{et al.}
%``High-Precision Measurement of the Proton Elastic Form Factor Ratio $\mu_pG_E/G_M$ at low $Q^2$,''
Phys. Lett. B \textbf{705}, 59-64 (2011).
%doi:10.1016/j.physletb.2011.10.002
%[arXiv:1102.0318 [nucl-ex]].
%241 citations counted in INSPIRE as of 21 Jun 2023

\bibitem{proton-EM-FF-Ex-2011-Ron}
%\cite{JeffersonLabHallA:2011yyi}
%\bibitem{JeffersonLabHallA:2011yyi}
G.~Ron \textit{et al.} [Jefferson Lab Hall A],
%``Low $Q^2$ measurements of the proton form factor ratio $mu_p G_E / G_M$,''
Phys. Rev. C \textbf{84}, 055204 (2011).
%doi:10.1103/PhysRevC.84.055204
%[arXiv:1103.5784 [nucl-ex]].
%128 citations counted in INSPIRE as of 21 Jun 2023
%\cite{BESIII:2019hdp}
%\bibitem{BESIII:2019hdp}
\bibitem{proton-EM-FF-Ex-2020-BESIII}
M.~Ablikim \textit{et al.} [BESIII],
%``Measurement of proton electromagnetic form factors in $e^+e^- \to p\bar{p}$ in the energy region 2.00 - 3.08 GeV,''
Phys. Rev. Lett. \textbf{124}, 042001 (2020).
%doi:10.1103/PhysRevLett.124.042001
%[arXiv:1905.09001 [hep-ex]].
%72 citations counted in INSPIRE as of 21 Jun 2023


\bibitem{proton-strange-FF-Ex-Muller-1997}
%\bibitem{SAMPLE}
B. Mueller {\it et al.}, Phys. Rev. Lett. {\bf 78}, 3824 (1997).

\bibitem{proton-strange-FF-Ex-Hasty-2000}
R.Hasty {\it et al.}, Science {\bf 290}, 2117 (2000).

\bibitem{proton-strange-FF-Ex-Spayde-2004}
D.T. Spayde {\it et al.}, Phys. Lett. {\bf B 583}, 79 (2004).


\bibitem{proton-strange-FF-Ex-HAPPEX}
K.A. Aniol {\it et al.} (HAPPEX), Phys. Rev. {\bf C 69}, 065501
(2004), Phys. Lett. {\bf B 635}, 275 (2006); A. Acha {\it et al.}
(HAPPEX), Phys. Rev. Lett. {\bf 98}, 032301 (2007).
%\bibitem{A4}

\bibitem{proton-strange-FF-Ex-A4}
F.E. Maas {\it et al.} (A4), Phys. Rev. Lett. {\bf 93}, 022002
(2004); Phys. Rev. Lett. {\bf 94}, 152001 (2005).

\bibitem{proton-strange-FF-Ex-G0}
D.S. Armstrong {\it et al.} (G0), Phys. Rev. Lett. {\bf 95}, 092001
(2005); C. Furget [G0], Nucl. Phys. Proc.
Suppl. {\bf 159}, 121 (2006).

\bibitem{Ex-Qweak}
%\cite{Qweak:2013zxf}
%\bibitem{Qweak:2013zxf}
D.~Androi\'c \textit{et al.} [Qweak],
%``First Determination of the Weak Charge of the Proton,''
Phys. Rev. Lett. \textbf{111}, 141803 (2013);
%doi:10.1103/PhysRevLett.111.141803
%[arXiv:1307.5275 [nucl-ex]].
D.~Androi\'c \textit{et al.} [Qweak],
%``Precision measurement of the weak charge of the proton,''
Nature \textbf{557}, 207 (2018).
%doi:10.1038/s41586-018-0096-0
%[arXiv:1905.08283 [nucl-ex]].;


\bibitem{proton-size-Ex-2010-Pohl}
%\bibitem{Pohl2010}
Pohl R {\it et al.}, Nature  {\bf 466}, 213 (2010).
%\bibitem{Antognini2013-Science}

\bibitem{proton-size-Ex-2013-Antognini}
Aldo Antognini {\it et al.}, Science {\bf 339}, 417 (2013).
%\cite{Fleurbaey:2018fih}

\bibitem{proton-size-Ex-2018-Fleurbaey}
%\bibitem{Fleurbaey:2018fih}
H.~Fleurbaey \textit{et al.},
%``New Measurement of the $1S-3S$ Transition Frequency of Hydrogen: Contribution to the Proton Charge Radius Puzzle,''
Phys. Rev. Lett. \textbf{120}, 183001 (2018).
%doi:10.1103/PhysRevLett.120.183001
%[arXiv:1801.08816 [physics.atom-ph]].

\bibitem{proton-size-Ex-2019-Bezginov}
%\cite{Bezginov:2019mdi}
%\bibitem{Bezginov:2019mdi}
N.~Bezginov {\it et al.},
%``A measurement of the atomic hydrogen Lamb shift and the proton charge radius,''
Science \textbf{365}, 1007 (2019);
%doi:10.1126/science.aau7807

\bibitem{proton-size-Ex-2019-Xiong}
%\cite{Xiong:2019umf}
%\bibitem{Xiong:2019umf}
W.~Xiong {\it et al.},
%``A small proton charge radius from an electron\textendash{}proton scattering experiment,''
Nature \textbf{575}, 147 (2019);
%doi:10.1038/s41586-019-1721-2

%Haiyan Gao, Marc Vanderhaeghen, Rev.Mod.Phys. 94 (2022) 1, 015002.

\bibitem{gamma-Z-exchange-Th-1984-Marciano}
W.~J.~Marciano and A. Sirlin, Phys.\ Rev.  D 27, 27 (1983);
{\bf 29}, 75 (1984); {\bf 31},  213(E) (1985);


\bibitem{gamma-Z-exchange-Th-2003-Erler}
%\cite{Erler:2003yk}
%\bibitem{Erler:2003yk}
J.~Erler, A.~Kurylov and M.~J.~Ramsey-Musolf,
%``The Weak charge of the proton and new physics,''
Phys. Rev. D \textbf{68}, 016006 (2003);
%doi:10.1103/PhysRevD.68.016006
%[arXiv:hep-ph/0302149 [hep-ph]].

\bibitem{gamma-Z-exchange-Th-2007-Zhou}
H. Q. Zhou, C. W. Kao and S. N. Yang, Phys. Rev. Lett
{\bf 99}, 262001 (2007); {\it ibid.} 100, 059903(E) (2008);

\bibitem{gamma-Z-exchange-Th-2008-Tjon}
J. A. Tjon and W. Melnitchouk, Phys. Rev. Lett. {\bf 100}, 082003 (2008).


\bibitem{gamma-Z-exchange-Th-DR-Gorchtein-2009}
%\cite{Gorchtein:2008px}
%\bibitem{Gorchtein:2008px}
M.~Gorchtein and C.~J.~Horowitz,
%``Dispersion gamma Z-box correction to the weak charge of the proton,''
Phys. Rev. Lett. \textbf{102}, 091806 (2009);
%doi:10.1103/PhysRevLett.102.091806
%[arXiv:0811.0614 [hep-ph]].
%80 citations counted in INSPIRE as of 21 Jun 2023


\bibitem{gamma-Z-exchange-Th-DR-Sibirtsev-2010}
%\cite{Sibirtsev:2010zg}
%\bibitem{Sibirtsev:2010zg}
A.~Sibirtsev, P.~G.~Blunden, W.~Melnitchouk and A.~W.~Thomas,
%``gamma-Z corrections to forward-angle parity-violating e-p scattering,''
Phys. Rev. D \textbf{82}, 013011 (2010).
%doi:10.1103/PhysRevD.82.013011
%[arXiv:1002.0740 [hep-ph]].

\bibitem{gamma-Z-exchange-Th-DR-2011-Blunden}
%\cite{Blunden:2011rd}
%\bibitem{Blunden:2011rd}
P.~G.~Blunden, W.~Melnitchouk and A.~W.~Thomas,
%``New formulation of gamma-Z box corrections to the weak charge of the proton,''
Phys. Rev. Lett. \textbf{107}, 081801 (2011);
%doi:10.1103/PhysRevLett.107.081801
%[arXiv:1102.5334 [hep-ph]].

\bibitem{gamma-Z-exchange-Th-DR-2011-Rislow}
%\cite{Rislow:2010vi}
%\bibitem{Rislow:2010vi}
B.~C.~Rislow and C.~E.~Carlson,
%``Contributions from $\gamma Z$ box diagrams to parity violating elastic $ep$ scattering,''
Phys. Rev. D \textbf{83}, 113007 (2011);
%doi:10.1103/PhysRevD.83.113007
%[arXiv:1011.2397 [hep-ph]].

\bibitem{gamma-Z-exchange-Th-DR-2011-Gorchtein}
%\cite{Gorchtein:2011mz}
%\bibitem{Gorchtein:2011mz}
M.~Gorchtein, C.~J.~Horowitz and M.~J.~Ramsey-Musolf,
%``Model-dependence of the $\gamma Z$ dispersion correction to the parity-violating asymmetry in elastic $ep$ scattering,''
Phys. Rev. C \textbf{84}, 015502 (2011);
%doi:10.1103/PhysRevC.84.015502
%[arXiv:1102.3910 [nucl-th]].

\bibitem{gamma-Z-exchange-Th-DR-2013-Hall}
%\cite{Hall:2013hta}
%\bibitem{Hall:2013hta}
N.~L.~Hall, P.~G.~Blunden, W.~Melnitchouk, A.~W.~Thomas and R.~D.~Young,
%``Constrained $\gamma Z$ interference corrections to parity-violating electron scattering,''
Phys. Rev. D \textbf{88}, 013011 (2013);
%doi:10.1103/PhysRevD.88.013011
%[arXiv:1304.7877 [nucl-th]].


\bibitem{gamma-Z-exchange-Th-DR-2019-Seng}
%\cite{Seng:2019plg}
%\bibitem{Seng:2019plg}
C.~Y.~Seng and U.~G.~Mei\ss{}ner,
%``Toward a First-Principles Calculation of Electroweak Box Diagrams,''
Phys. Rev. Lett. \textbf{122}, 211802 (2019).
%doi:10.1103/PhysRevLett.122.211802
%[arXiv:1903.07969 [hep-ph]].
%29 citations counted in INSPIRE as of 21 Jun 2023





\bibitem{gamma-Z-exchange-low-erergy-2023-zhouhq}
Qian-Qian Guo and Hai-Qing Zhou,  Phys. Rev. C \textbf{108}, 035501 (2023).


\bibitem{Gorchtein-2011-PRC}
M.~Gorchtein, C.~J.~Horowitz and M.~J.~Ramsey-Musolf,
%``Model-dependence of the $\gamma Z$ dispersion correction to the parity-violating asymmetry in elastic $ep$ scattering,''
Phys. Rev. C \textbf{84}, 015502 (2011).
%doi:10.1103/PhysRevC.84.015502
%[arXiv:1102.3910 [nucl-th]].
%103 citations counted in INSPIRE as of 07 Oct 2023


%%%%%%%%%%%%%%%%%%%%%%%%%%%%%%%%%%%%%%%%%%%%%%%%%%%%%%%%%%%%%%%%%
\bibitem{FenyCalc}
R. Mertig, M. Bohm and A. Denner, Comput. Phys. Commun. {\bf 64}, 345 (1991);
V. Shtabovenko, R. Mertig and F. Orellana, Comput. Phys. Commun. {\bf 207}, 432 (2016).

%%%%%%%%%%%%%%%%%%%%%%%%%%%%%%%%%%%%%%%%%%%%%%%%%%%%%%%%%%%%%%%%%
\bibitem{PacakgeX}
H. H. Patel,  Comput. Phys. Commun. {\bf 197}, 276-290 (2015);
H. H. Patel,  Comput. Phys. Commun. {\bf 218}, 66-70 (2017).


%%%%%%%%%%%%%%%%%%%%%%%%%%%%%%%%%%%%%%%%%%%%%%%%%%%%%%%%%%%%%%%%%
\bibitem{LoopTools}
T. Hahn, M. Perez-Victoria, Comput. Phys. Commun. {\bf 118}, 153 (1999).

%\cite{Becker:2018ggl}
\bibitem{P2-experiment-2018}
D.~Becker,  {\it et al.},
%``The P2 experiment,''
Eur. Phys. J. A \textbf{54}, 208 (2018).
%doi:10.1140/epja/i2018-12611-6
%[arXiv:1802.04759 [nucl-ex]].




\end{thebibliography}
\end{document}